\documentclass[aps,preprint]{revtex4}%
\usepackage{amsfonts}
\usepackage{amsmath}
\usepackage{amssymb}
\usepackage{graphicx}%
\usepackage{mathtools}
\providecommand{\U}[1]{\protect\rule{.1in}{.1in}}

\begin{document}
\title{Spatiotemporal dissipative solitons and vortices in a
multi-transverse-mode fiber laser}
\author{Thawatchai Mayteevarunyoo$^{1}$, Boris A. Malomed$^{2,3}$, and Dmitry V.
Skryabin$^{3,4}$}
\affiliation{$^{1}$Department of Electrical and Computer Engineering, Faculty of
Engineering, Naresuan University, Phitsanulok 65000, Thailand}
\affiliation{$^{2}$Department of Physical Electronics, School of Electrical Engineering,
Faculty of Engineering, and Center for Light-Matter Interaction, Tel Aviv
University, Tel Aviv 69978, Israel}
\affiliation{$^{3}$ITMO University, St. Petersburg 197101, Russia}
\affiliation{$^{4}$Department of Physics, University of Bath, Bath, BA2 7AY, UK}

\begin{abstract}
We introduce a model for spatiotemporal modelocking in multimode fiber
lasers, which is based on the (3+1)-dimensional cubic-quintic complex
Ginzburg-Landau equation (cGLE) with conservative and dissipative
nonlinearities and a 2-dimensional transverse trapping potential. Systematic
numerical analysis reveals a variety of stable nonlinear modes, including
stable fundamental solitons and breathers, as well as solitary vortices with
winding number $n=1$, while vortices with $n=2$ are unstable, splitting into
persistently rotating bound states of two unitary vortices. A characteristic
feature of the system is bistability between the fundamental and vortex
spatiotemporal solitons.
\end{abstract}
\maketitle


\section{Introduction and the model}

New physical principles for generating and controlling ultrashort pulses are
under constant scrutiny of the scientific community. A recent prominent topic
of the research in this area is complex multi-mode spatiotemporal dynamics in
optical fibers used for both generation and delivery of short pulses, see,
e.g., \cite{grelu,wise1,lush,jena} for review. As intensity of light is
increased through tighter focusing in both time and space, it starts
interacting with matter nonlinearly, which is crucially important for the
generation of ultrashort pulses and their spectral and temporal shaping. While
the structure of the short pulses is intrinsically multimode (in a cavity) and
multi-wavelength in the propagation direction (in free space), their
transverse multimode structure has been traditionally considered as a thing to
avoid and a special care is commonly taken to ensure that a pulse is carried
by a single transverse mode. Nevertheless, transversely multimode ultrashort
pulses have recently become a topic of intense research. Their spatial
complexity is a feature to be controlled and utilized \cite{wise1},
fiber-based devices being the most prominent setting in this context. In
particular, multi-transverse-mode passive fibers have been used for boosting
transmission capacity of communication lines by means of spatial-division
multiplexing \cite{sdm}. Strongly nonlinear propagation of pulses in
graded-index (GRIN) fibers with a transversely parabolic refractive index has
been investigated in the context of the beam-profile self-cleaning, as well as
the generation of continuum radiation and solitons
\cite{grin1,grin2,grin_a,grin_b,grin_c,sol1,sol2,r31,sol3}. Spatiotemporal
mode-locking (STML) of longitudinal and transverse modes in a fiber-laser
system was recently reported in \cite{exp1}, with follow-up results
contributing to this area. Those include observations of solitons and soliton
bound states \cite{exp2,exp3}, a proposal for the multi-transverse-mode
gain-saturation mechanism \cite{exp4}, and direct generation of vortex beams
\cite{exp5}. The latter result is directly relevant and motivational for the
present work, though the above-mentioned recent STML results predominantly
dealt with the normal-GVD (group-velocity-dispersion) case, while we are
address the case of anomalous GVD in this work. Previously, this case has been
chiefly considered in studies with suppressed transverse effects \cite{grelu}.
Multimode fibers are closely related to multi-core ones, since coupled cores
introduce multiple transverse modes, whose number is equal the numbers of
cores. Pulse compression, light bullets and spatiotemporal discrete vortices
all have been studied in the multi-core fibers \cite{turyt,jena1,jena2}. STML
in various multi-core waveguide systems has also been addressed by several
groups \cite{winf,kutz,egg}. It is relevant to mention that studies of STML
and soliton effects in (non-fiber) laser cavities had started decades
ago---see, e.g., \cite{auston,smith,firth}.

Mode-locking effects and formation of dissipative solitons have been
traditionally (and successfully) modeled using various realizations of the
complex Ginzburg-Landau equation (cGLE), its ubiquitous version being one with
the cubic-quintic nonlinearity \cite{grelu,book}. The quintic term supports
the subcritical transition to the lasing state, thereby providing conditions
for mode-locking and the creation of stable solitons at subcritical pumping
levels \cite{book}. Locking of the longitudinal modes in
single-transverse-mode fiber lasers is described by (1+1)-dimensional cGLE,
while the consideration of multi-mode fibers opens the way to the realm of the
(3+1)D cGLE. The lateral confinement of the fiber modes immediately requires
inclusion of the trapping potential acting in two transverse dimensions, which
is a basic parabolic potential for GRIN fibers. This potential is a feature
creating a new playground relative to the quite extensive studies of
\textquotedblleft light bullets" \cite{agr,bul1,bul2,bul3} and vortex solitons
\cite{vort1,vort2,vort3,vort4,vort5} in (3+1)D cGLE. Below we use the (3+1)D
cGLE including conservative and dissipative cubic-quintic terms and a 2D
parabolic potential.

Measuring the propagation distance along the fiber, beam's width, and time in
units of suitably defined scales $L$, $w$ and $\tau$, the cGLE is written as
\begin{equation}%
\begin{multlined}
i\frac{\partial\psi}{\partial z}+\frac{L}{2\tau^{2}}k^{\prime\prime}\left(
1-i\beta\right)  \frac{\partial^{2}\psi}{\partial{t}^{2}}+\frac{L}{2k_{0}%
w^{2}}\left(  \frac{\partial^{2}}{\partial{x}^{2}}+\frac{\partial^{2}%
}{\partial{y}^{2}}\right)  \psi-\frac{k_{0}gw^{2}L}{2}\left(  {x}^{2}+{y}%
^{2}\right)\psi \\
+iL\varepsilon^{\prime}\psi+L\left(\eta^{\prime}-i\alpha^{\prime}\right)|\psi|^{2}\psi
+L\left(\nu^{\prime}+i\mu^{\prime}\right)  |\psi|^{4}\psi=0.
\end{multlined}
\label{3DcGLE_0}%
\end{equation}
Here $z$, $t$ and $x,y$ are the dimensionless distance, time, and transverse
coordinate, and time is defined in the reference frame moving with the pulse's
group velocity. Further, $k^{\prime\prime}$ is the GVD coefficient, $\beta$
the spectral-filtering coefficient, $k_{0}$ the propagation constant of the
fundamental mode, $g$ the GRIN parameter, and $\varepsilon\equiv
L\varepsilon^{\prime}>0$ the dimensionless linear loss over length $L$. If
$\psi$ is the dimensionless field envelope, then the scaled (dimensionless)
parameters $\eta=L\eta^{\prime}$, $\alpha=L\alpha^{\prime}>0$, $\nu
=L\nu^{\prime}$ and $\mu=L\mu^{\prime}>0$ determine ratios between $L$ and the
nonlinear lengths associated with the Kerr effect, cubic gain, quintic
nonlinearity, and nonlinear saturation, respectively. We now fix
$L=1/(k_{0}gw^{2})$, $w^{2}=L/k_{0}$ and $\tau^{2}=Lk^{\prime\prime}$ to write
the spatiotemporal cGLE (\ref{3DcGLE_0}) with the cubic-quintic terms in the
scaled form:
\begin{equation}
i\frac{\partial\psi}{\partial z}+\frac{1}{2}\left(  1-i\beta\right)
\frac{\partial^{2}\psi}{\partial t^{2}}+\frac{1}{2}\left(  \frac{\partial^{2}%
}{\partial x^{2}}+\frac{\partial^{2}}{\partial y^{2}}\right)  \psi-\left(
x^{2}+y^{2}-i\varepsilon\right)  \psi+\left(  \eta-i\alpha\right)  |\psi
|^{2}\psi+\left(  \nu+i\mu\right)  |\psi|^{4}\psi=0.\label{3DcGLE}%
\end{equation}
Nontrivial solutions can be supported by the cubic gain in the competition
with the linear and quintic losses, if the gain coefficient exceeds a
threshold value \cite{May2}%
\begin{equation}
\alpha=A\alpha_{\mathrm{thr}},~\alpha_{\mathrm{thr}}\equiv\left(
2\sqrt{\varepsilon\mu}\right)  \label{gain_thr}%
\end{equation}
with $A>1$. We here fix values $A=1.5$ and $\beta=\nu=0$, which adequately
represent the generic case, while the linear loss and nonlinear gain,
$\varepsilon$ and $\mu$, are varied.

Thus, in this work we focus on predicting STML scenarios in the setup based on
the interplay of the cubic-quintic dissipative nonlinearity with the anomalous
GVD, transverse diffraction, parabolic GRIN potential, and Kerr self-focusing,
in the absence of temporal and spatial filtering. In fact, the prediction of
patterns which are stable without the help of the filtering is an essential
novelty of the analysis.

\section{Numerical results}

We look for stationary localized solutions, in the form of $\psi
(x,y,t,z)=\phi(x,y,t)\exp\left(  iqz\right)  $, where $q$ is a real
propagation constant, and function $\phi(x,y,t)$ satisfies the equation
following from the substitution of this ansatz in Eq. (\ref{3DcGLE}) with
$\beta=\nu=0$:
\begin{equation}
-q\phi+\frac{1}{2}\left(  \frac{\partial^{2}}{\partial t^{2}}+\frac
{\partial^{2}}{\partial x^{2}}+\frac{\partial^{2}}{\partial y^{2}}\right)
\phi-\left(  x^{2}+y^{2}-i\varepsilon\right)  \phi+\left(  \eta-i\alpha
\right)  |\phi|^{2}\phi+i\mu|\phi|^{4}\phi=0.\label{phi_solution}%
\end{equation}
Splitting complex equation (\ref{phi_solution}) into real and imaginary parts,
solutions for localized states have been produced by means of the modified
squared-operator method \cite{yang}. The solutions are characterized by the
value of their total energy, i.e., integral norm,%
\begin{equation}
P=\int\int\int\left\vert \psi(x,y,t)\right\vert ^{2}dxdydt.\label{Power}%
\end{equation}

\subsection{Fundamental dissipative solitons}

\begin{figure}[ptbh]
\centering\includegraphics[width=4in]{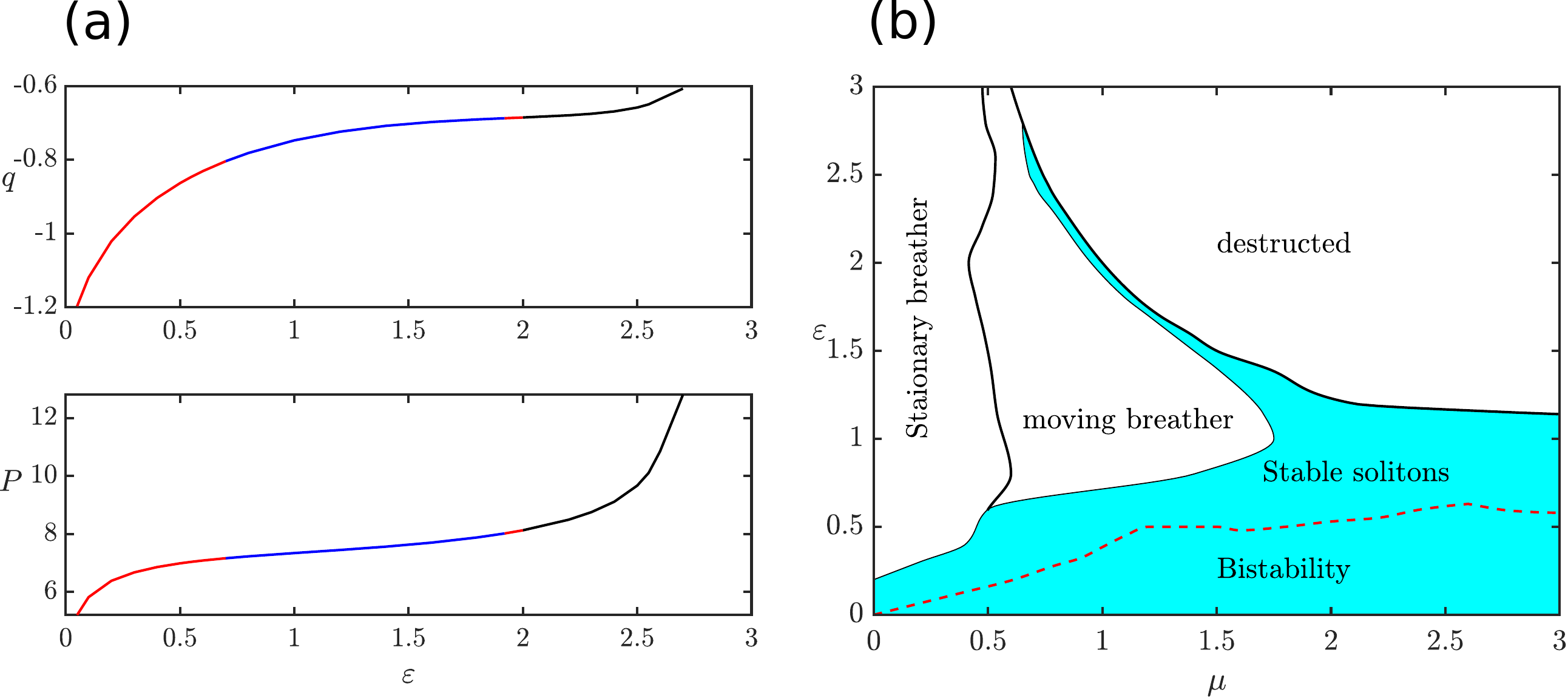}\caption{(a) Numerically
computed propagation constant $q$ and integral power $P$ [see Eq.
(\ref{Power})] of fundamental three-dimensional solitons versus $\varepsilon$,
for fixed $\mu=1$ in Eq. (\ref{3DcGLE}). Red, blue and black segments
correspond to stable fundamental solitons, moving breathers, and eventually
destructed modes, respectively. (b) The stability chart for the stationary
fundamental solitons and modes into which unstable solitons are spontaneously
transformed. Indicated inside the stability area of solitons is the
bistability region, in which solitary vortices with winding numbers are stable
too, cf. Fig. 5(b). }%
\label{fig1}%
\end{figure}

\begin{figure}[ptbh]
\centering\includegraphics[width=4in]{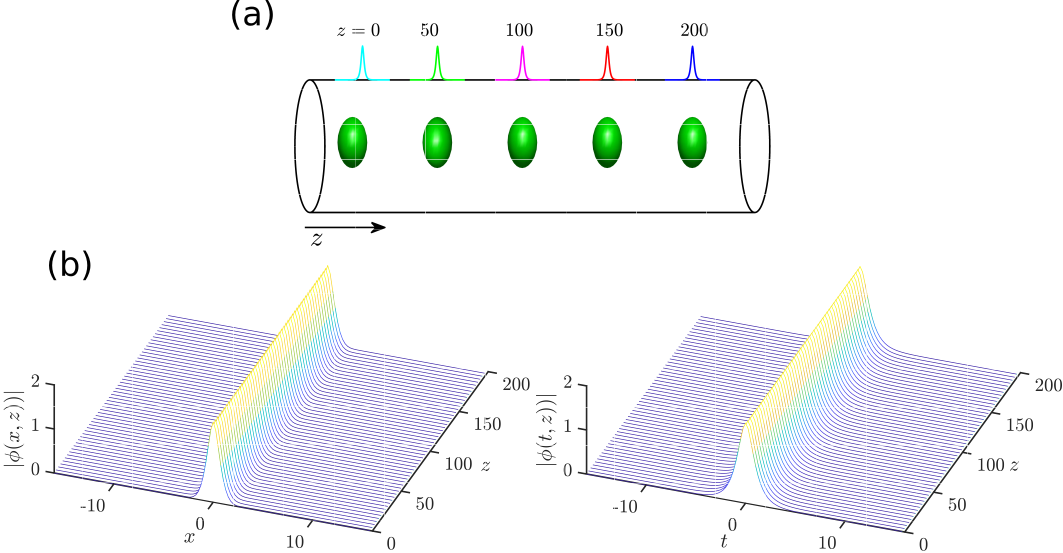}\caption{(a) The evolution of a
stable fundamental soliton, for $\mu=1,\varepsilon=0.5$, with
$q=-0.8635,P=6.9898$, is displayed by means of isosurfaces of $\left\vert
\psi\left(  x,y,z,t\right)  \right\vert $. (b) Left and right panels display
spatial and temporal cross sections of the soliton's shape, $|\psi(x,0,z,0)|$
and $|\psi(0,0,z,t)|$, respectively.}%
\label{fig2}%
\end{figure}

To determine numerical solutions and the propagation constant simultaneously,
the Gaussian input was used, $\phi_{0}=1.5~\mathrm{sech}\left(  x^{2}%
+y^{2}\right)  \mathrm{\exp}(-1/2t^{2})$. The existence and stability of
numerically found fundamental (zero-vorticity) dissipative solitons, produced
by Eq. (\ref{phi_solution}), is summarized in Fig. \ref{fig1}(a), which
displays the numerically found $q$ and $P$ (integral power) vs. the
linear-loss parameter, $\varepsilon$, for a fixed value of the quintic-loss
coefficient, $\mu=1$, while the cubic gain is fixed as per Eq. (\ref{gain_thr}%
). The stability was identified by means of systematic simulations of
perturbed evolution of the solitons in the framework of Eq. (\ref{3DcGLE}).
Figure \ref{fig1}(b) summarizes the results by plotting stability boundaries
for the dissipative solitons in the plane of $\left(  \mu,\varepsilon\right)
$. The shape and evolution of the stable solitons are illustrated by Fig.
\ref{fig2}. 

\begin{figure}[ptbh]
\centering\includegraphics[width=4in]{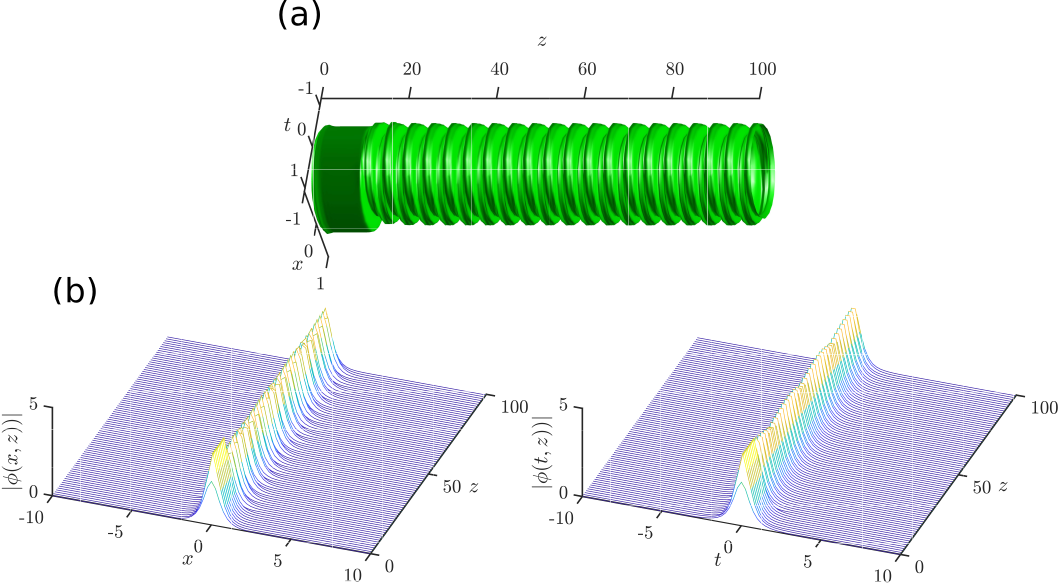}\caption{(a) The evolution of a
robust quiescent breather, into which an unstable fundamental soliton
transforms for $\mu=0.3,\varepsilon=2.8$, which corresponding to
$q=-2.0446,P=5.3081$, shown by means of the isosurface of $\left\vert
\psi\left(  x,y,z,t\right)  \right\vert $. (b) Left and right panels show the
evolution in the spatial and temporal of cross sections, $|\psi(x,0,z,0)|$ and
$|\psi(0,0,z,t)|$, respectively. }%
\label{fig3}%
\end{figure}

\begin{figure}[ptbh]
\centering\includegraphics[width=4in]{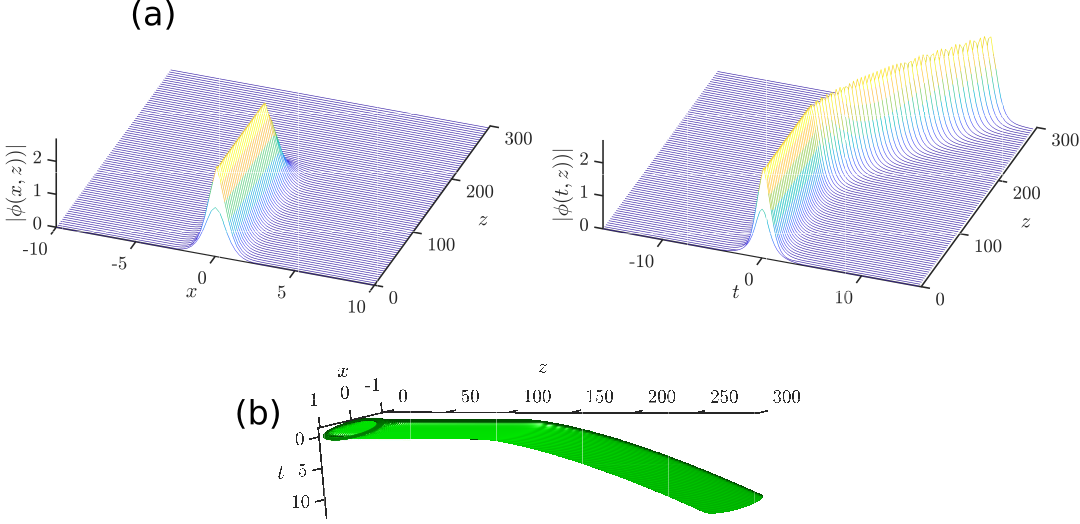}\caption{A robust moving
breather for $\mu=1.0,\varepsilon=1.8$, into which the respective unstable
soliton, with $q=-1.7927,P=4.5920$, spontaneously transforms. (a) The
evolution in the spatial and temporal cross sections, $|\psi(x,0,z,0)|$ (left)
and $|\psi(0,0,z,t)|$ (right). (b) The spatiotemporal dynamics with temporal
motion.}%
\label{fig4}%
\end{figure}

Weakly unstable fundamental solitons can be easily found too. They
spontaneously transform into quiescent or moving breathers, see examples in
Fig. \ref{fig3} and \ref{fig4}, respectively. The breathers are robust
dynamical modes, which keep integrity in the course of indefinitely long
subsequent evolution. Their stability regions in the parameter plane of
$\left(  \mu,\varepsilon\right)  $ are also drawn in Fig. \ref{fig1}(b). Note,
in particular, that the region of destruction of all modes (i.e., onset of
spatial \textquotedblleft turbulence\textquotedblright\ as a result of the
simulated evolution) occupies, quite naturally, an area in which the loss and
gain coefficients [recall that the latter one is taken as per Eq.
(\ref{gain_thr})] are too large to maintain the stability of regular modes,
such as solitons and breathers. The same peculiarity is observed in stability
charts displayed below for vortex states in Figs. \ref{fig5}(b) and
\ref{fig10}(b). 

\subsection{Vortices with winding number $n=1$}

\begin{figure}[ptbh]
\centering\includegraphics[width=4in]{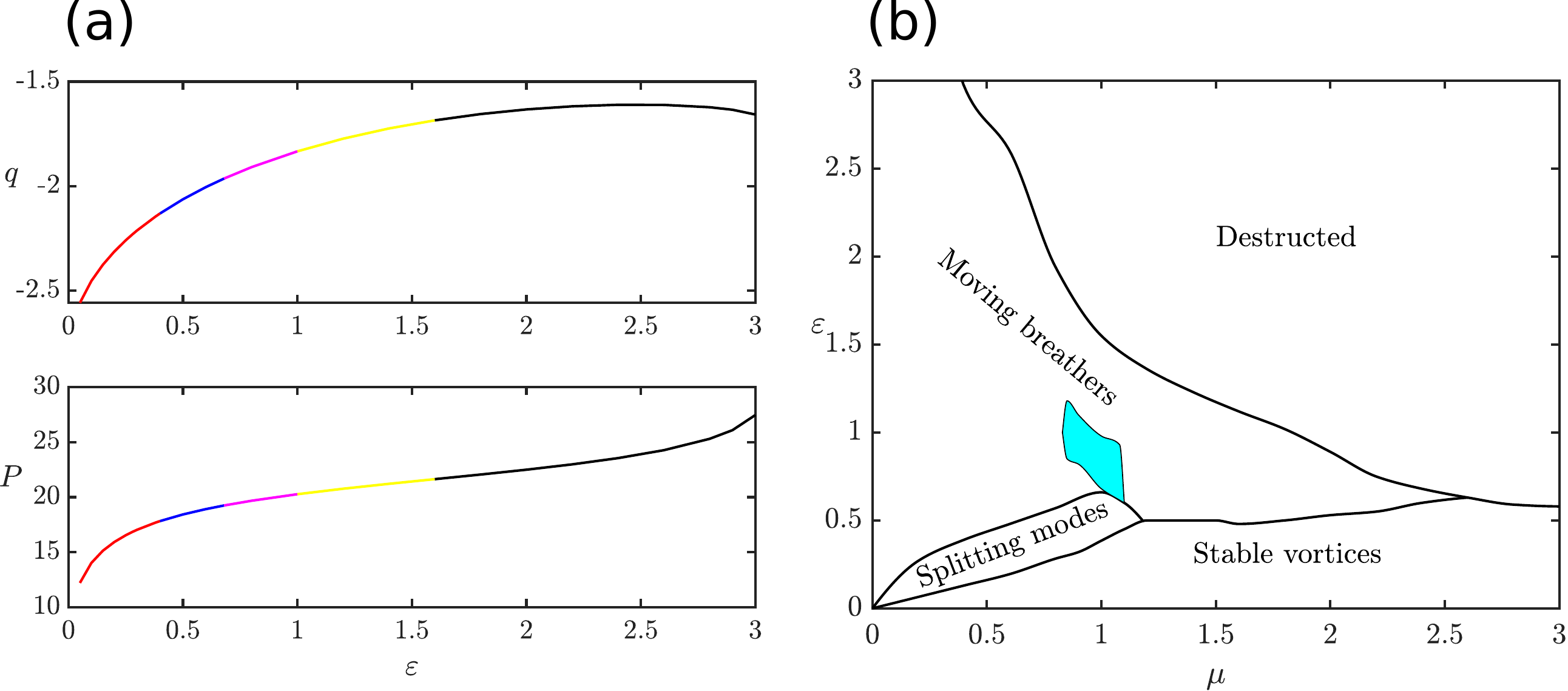}\caption{(a) The numerically
found values of $q$ (top panel) and $P$ (bottom panel) for stationary vortices
with winding number $n=1$ versus $\varepsilon$ for fixed $\mu=1.0$. Red, blue,
magenta, yellow and black segments correspond to stable vortices, splitting
into fundamental solitons, chaotically oscillating modes, moving vortex
breathers, and destruction of unstable vortices, respectively. (b) The
stability chart for vortices with $n=1$ in the plane of $\left(
\mu,\varepsilon\right)  $. The central shaded area in (b) is populated by
chaotically oscillating localized modes, see an example in Fig. \ref{fig8}.
Note that the stability area of vortices is completely covered by the region
in which the fundamental solitons are stable [cf. Fig. \ref{fig1}(b)], i.e.,
it is a bistability area.}%
\label{fig5}%
\end{figure}

\begin{figure}[ptbh]
\centering\includegraphics[width=4in]{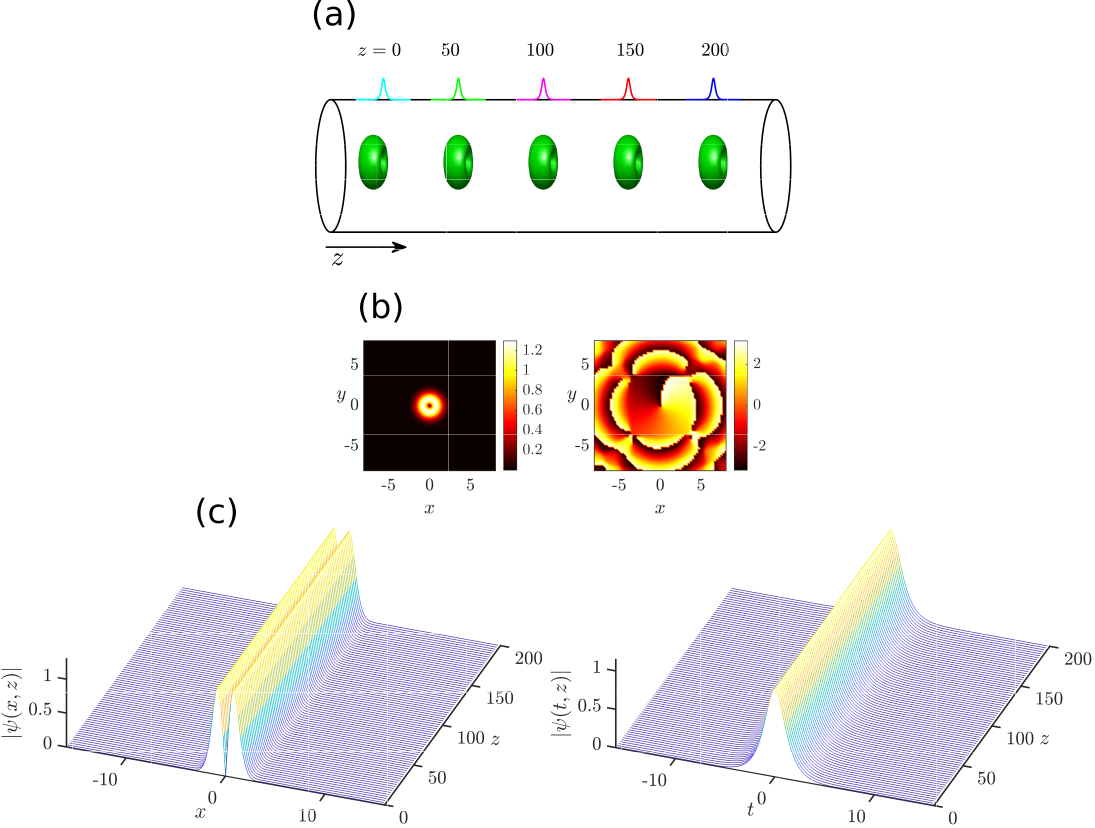}\caption{(a) The isosurface
evolution of a stable vortex found at $\mu=1.0,\varepsilon=0.2$, which
corresponding to $q=-2.3130,P=15.9142$. (b) Its amplitude and phase patterns
(c) Left and right panels display the evolution of the spatial and temporal
cross sections $|\psi(x,0,z,0)|$ and $|\psi(0,0,z,t)|$, respectively.}%
\label{fig6}%
\end{figure}

\begin{figure}[ptbh]
\centering\includegraphics[width=4in]{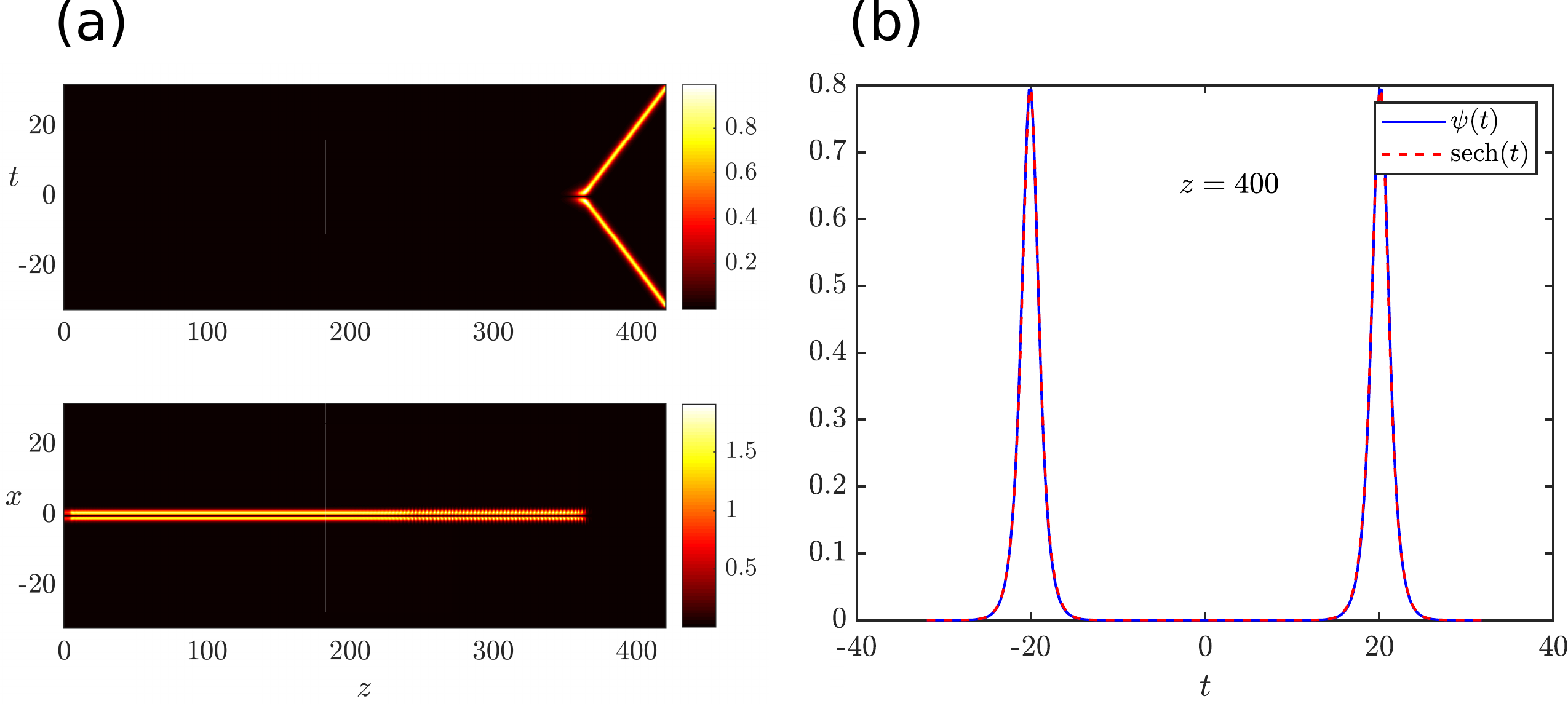}\caption{The spontaneous
splitting of an unstable vortex with $n=1$ into two fundamental solitons, at
$\mu=1.0,\varepsilon=0.6$. (a) The evolution of $|\psi|$ is shown in the $t$-
and $x$-cross sections (top and bottom panels, respectively). (b) The temporal
shape of the moving fundamental solitons, produced by the splitting, at
$z=400$\ (the blue line), fitted to the standard soliton's shape,
$\mathrm{sech}(t)$.}%
\label{fig7}%
\end{figure}

\begin{figure}[ptbh]
\centering\includegraphics[width=4in]{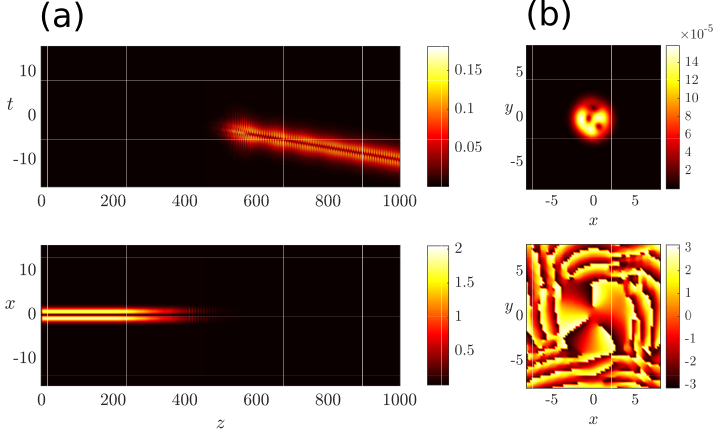}\caption{(a) The evolution of
$|\psi\left(  0,0,z,t\right)  |$ and $|\psi\left(  x,0,z,0\right)  |$ in the
$t$- and $x$- cross-sections, illustrating the spontaneous transformation of
an unstable stationary vortex with $n=1$ into a chaotically oscillating mode
at $\mu=1.0,\varepsilon=0.9$. (b) The respective amplitude and phase patterns
at $z=1000$ (eventual values of the amplitude are small, as a result of the
evolution). }%
\label{fig8}%
\end{figure}

\begin{figure}[ptbh]
\centering\includegraphics[width=4in]{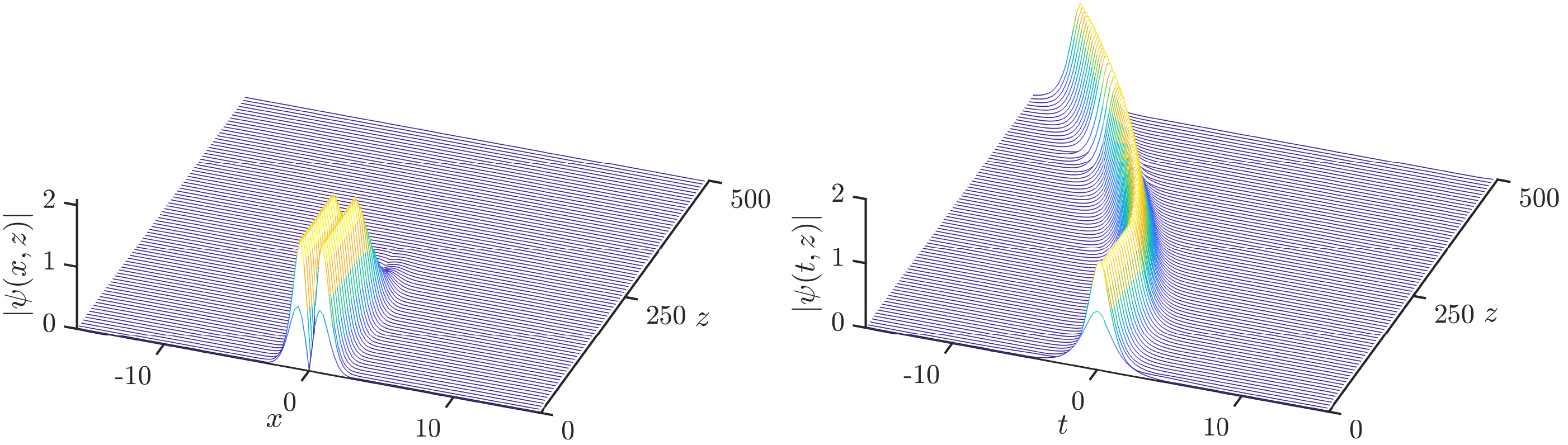}\caption{A moving breather,
into which an unstable stationary vortex, with $q=-2.9276,P=9.9546$, is
spontaneously transformed (keeping its vorticity) at $\mu=1.0,\varepsilon
=1.2$. The evolution of cross sections $|\psi\left(  x,0,z,0\right)  |$ and
$|\psi\left(  0,0,z,t\right)  |$ is displayed in the left and right panels,
respectively.}%
\label{fig9}%
\end{figure}

\begin{figure}[ptbh]
\centering\includegraphics[width=3in]{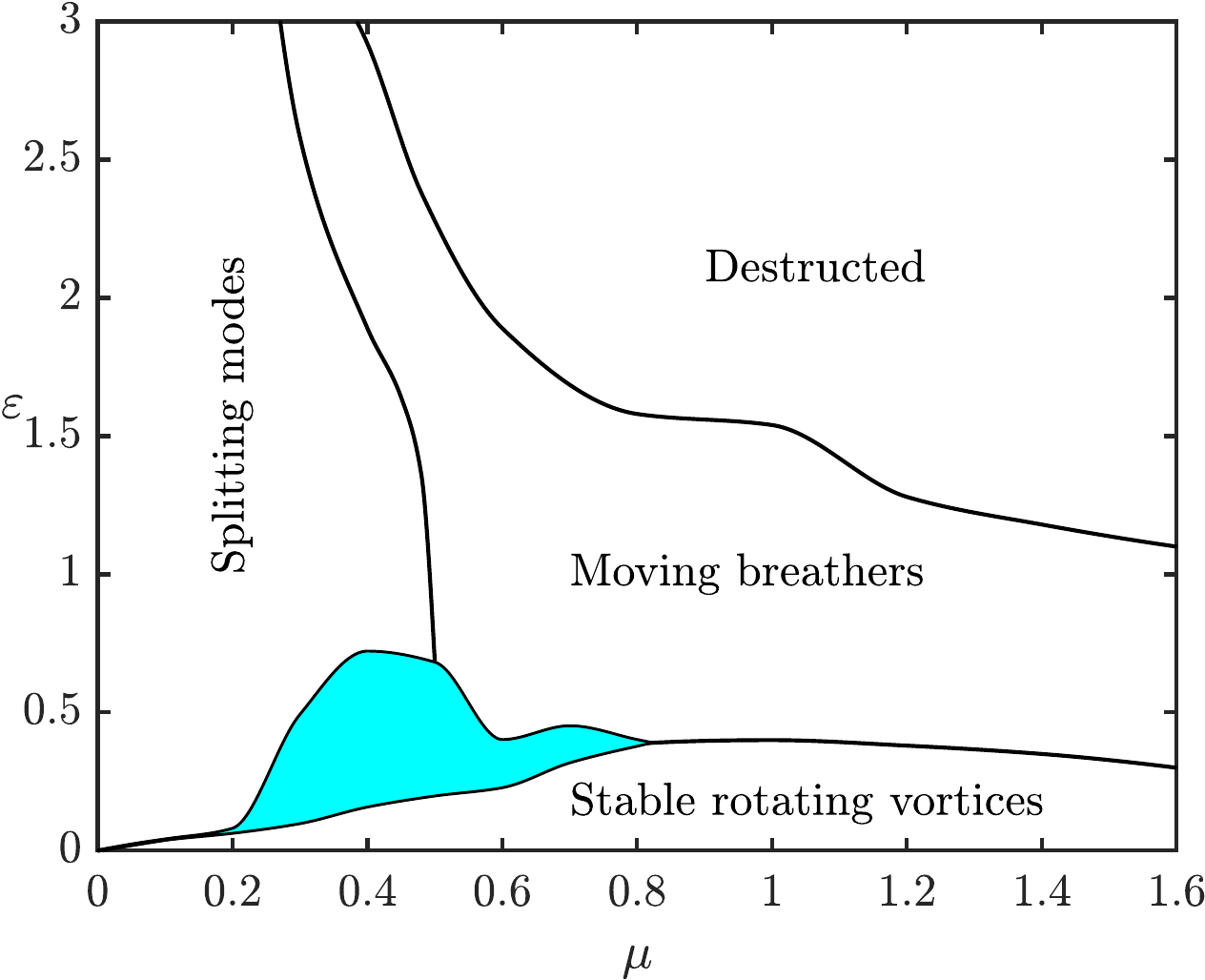}\caption{The stability chart
in the plane of $\left(  \mu,\varepsilon\right)  $ for vortices with $n=2$.
The shaded area corresponds to the \textquotedblleft wobbling
breathers\textquotedblright. }%
\label{fig10}%
\end{figure}

Localized vortex states with winding number (topological charge) $n=1$ were
constructed as solutions of Eq. (\ref{phi_solution}), starting with input,
written in polar coordinates $\left(  r,\theta\right)  $ (in the plane of
$\left(  x,y\right)  $) as $\phi_{0}=2.5r~\mathrm{sech}\left(  r^{2}\right)
~\mathrm{sech}\left(  t\right)  \exp\left(  i\theta\right)  $. As a result,
two coexisting vortex families, one unstable and one stable, have been found.
Figure \ref{fig5}(a) displays the respective numerically found values of $q$
and integral power $P$ versus $\varepsilon$. In particular, the stable (red)
branch was found in the interval of $0<\varepsilon<0.4$ for $\mu=1.0$.
Unstable vortex states split into fundamental solitons (in the blue segment),
or transform into chaotically oscillating modes (magenta) or moving breathers,
which keep the initial vorticity (yellow), or suffer destruction (black). The
results are summarized in Fig. \ref{fig5}(b) in terms of the $\left(
\mu,\varepsilon\right)  $ plane, where the shaded area pertains to chaotically
oscillating modes. Typical examples of the stable vortices, splitting,
chaotically oscillating modes, and moving vortex breathers are shown in Fig.
\ref{fig6}, \ref{fig7}, \ref{fig8} and \ref{fig9}, respectively.




An obviously important conclusion, demonstrated by the comparison of Figs.
\ref{fig5}(b) and \ref{fig1}(b), is the bistability between the fundamental
solitons and solitary vortices, as the solitons' stability area completely
covers that of the vortices. In fact, this feature may be interpreted as
tri-stability, as stable vortices may have winding numbers $+1$ and $-1$. This
fact may find applications to optical switching in the multimode fibers, and
is of considerable interest in terms of general properties of vortex solitons
in dissipative media. The switching may be controlled by adjusting the power:
it is seen in the bottom panels of Figs. \ref{fig1}(a) and \ref{fig5}(a) that
the power of a stable vortex is $\simeq3$ times larger than the power of a
coexisting stable fundamental soliton. \begin{figure}[ptbh]
\centering\includegraphics[width=4in]{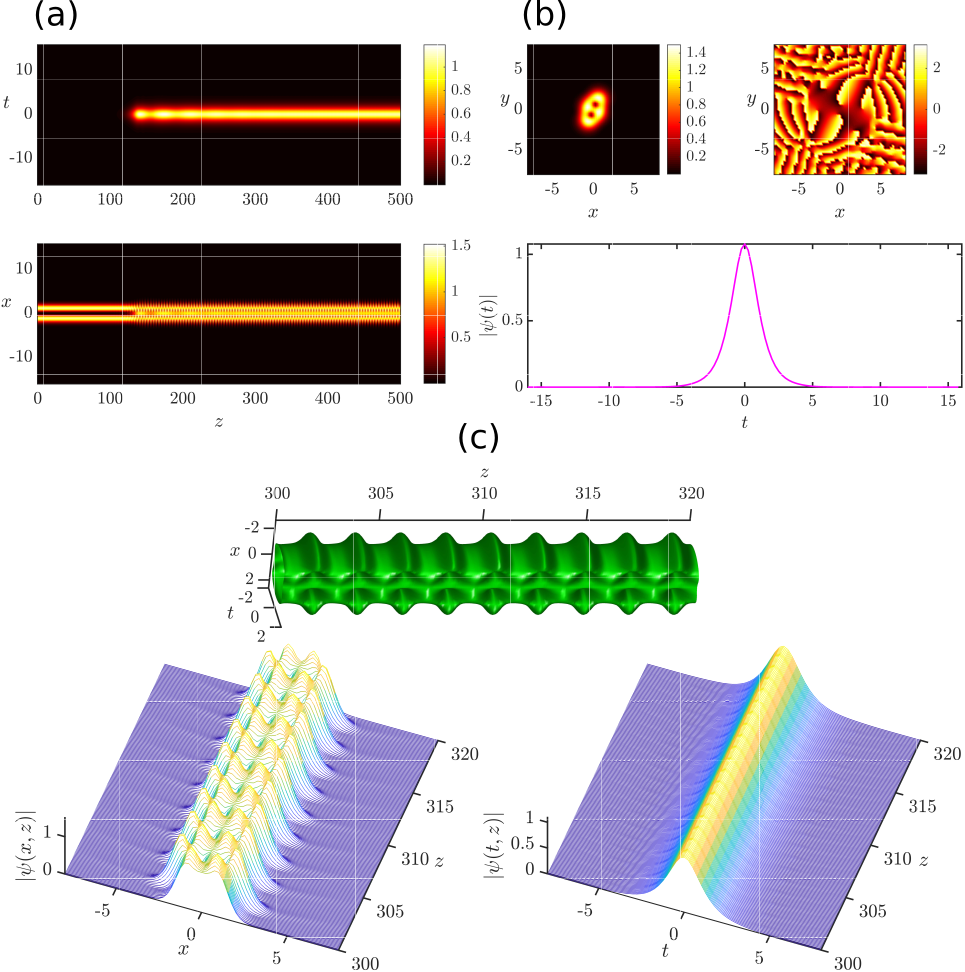}\caption{{}Spontaneous
splitting of an unstable vortex with $n=2$ into a stably rotating bound state
of two vortices with $n=1$ at $\mu=0.4$, $\varepsilon=0.1$ (a) The evolution
of the field in the temporal and spatial cross sections, $|\psi(0,0,z,t)|$ and
$|\psi(x,0,z,0)|$. (b) Amplitude and phase patterns of $\psi(x,y)$ in the
plane of $t=0$ and (bottom) the temporal profile at the central point,
$x=y=0$, plotted at $z=500$ (c) The top, left and right panels zoom in on the
evolution of the absolute value of the field, and spatial and temporal cross
sections, respectively, in the interval of $300<z<320$. The established
evolution (rotation) period is $T_{z}=2.2$.}%
\label{fig11}%
\end{figure}\begin{figure}[tbh]
\centering\includegraphics[width=4in]{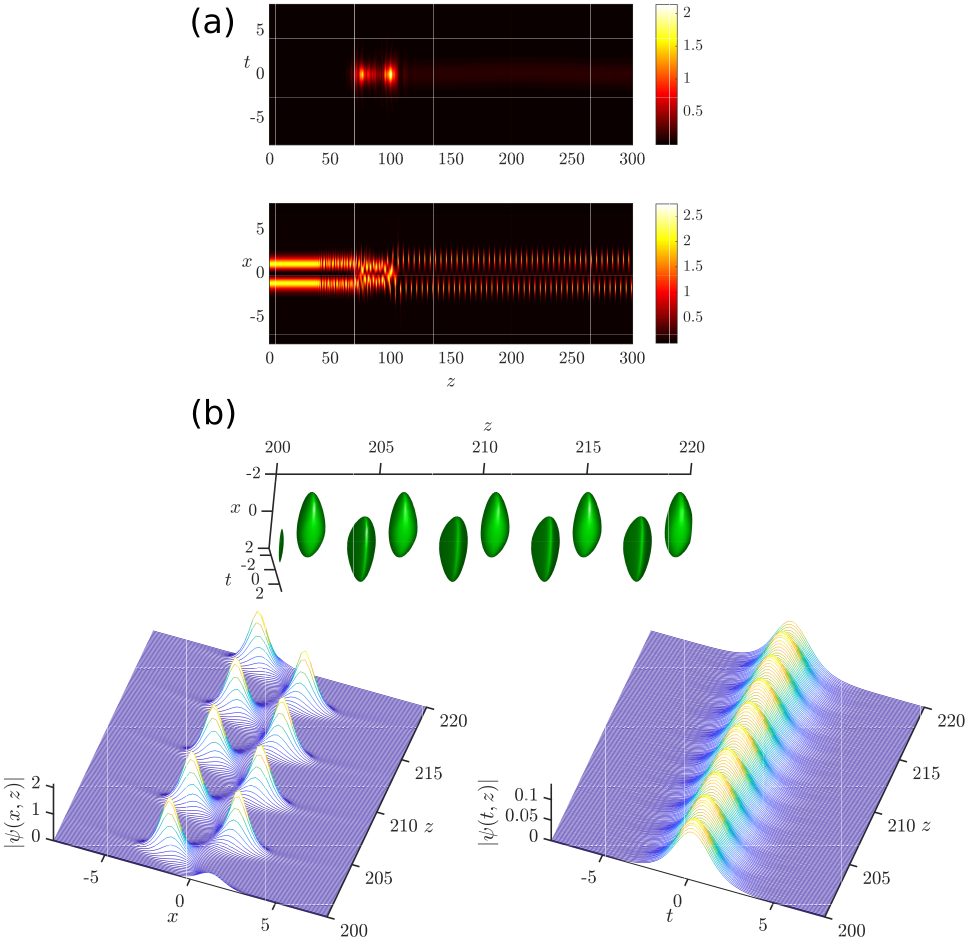}\caption{{}The
\textquotedblleft wobbling breather\textquotedblright\ at $\mu=0.4$,
$\varepsilon=0.6$ (a) The evolution of the field in the temporal and spatial
cross sections, $|\psi(0,0,z,t)|$ and $|\psi(x,0,z,0)|$. (b) The top, left and
right panels zoom in on the evolution of the absolute value of the field, and
spatial and temporal cross sections in the interval of $z=200-220$,
respectively. The established evolution period is $T_{z}=4.5$.}%
\label{fig12}%
\end{figure}\begin{figure}[tbh]
\centering\includegraphics[width=4in]{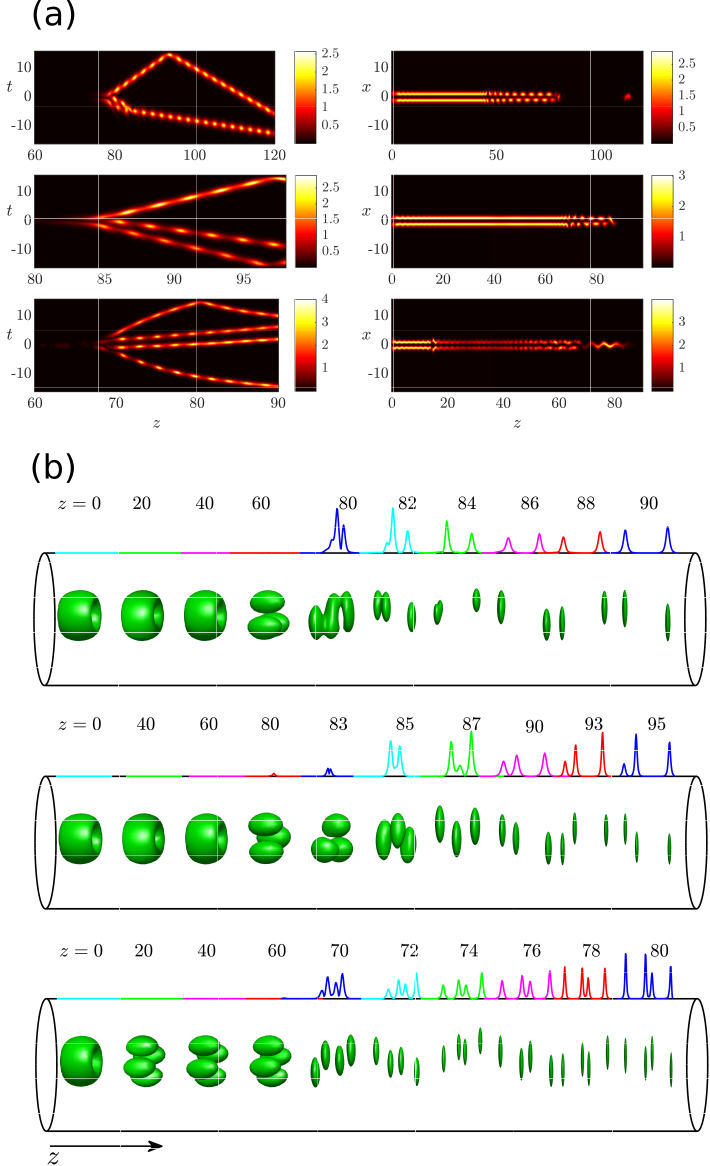}\caption{(a) {}Splitting of an
unstable breather with $n=2$ into $2$, $3$ and $4$ secondary breathers with
zero intrinsic vorticity at $\mu=0.4,\varepsilon=0.9$ (top), $\mu
=0.4,\varepsilon=1.4$ (middle), and $\mu=0.2,\varepsilon=3.0$ (bottom),
displayed by means of temporal cross sections, $|\psi\left(  0,0,z,t\right)
|$. Panels in the right column display the respective evolution in the spatial
cross section, $|\psi\left(  x,0,z,0\right)  |$. (b) The same as (a) but
illustrating the isosurface evolution in 3D.}%
\label{fig13}%
\end{figure}

\subsection{Vortices with $n=2$}

Double vortices, with winding number $n=2$, were constructed as solutions of
Eq. (\ref{phi_solution}), starting with input $\phi_{0}=2.5r~\mathrm{sech}%
\left(  r^{2}\right)  ~\mathrm{sech}\left(  t\right)  \exp\left(
2i\theta\right)  $. Systematic simulations have demonstrated that no double
vortices are stable. They spontaneously develop into various modes, which are
robust in different areas in the plane of $\left(  \mu,\varepsilon\right)  $,
as shown in Fig. \ref{fig10}. These are rotating bound states of two unitary
vortices, into which the double vortex splits, which is shown in Fig.
\ref{fig11}; \textquotedblleft wobbling breathers\textquotedblright, which
lose the original vorticity and perform spontaneous shuttle motion in the $x$
direction, coupled to intrinsic vibrations, see Fig. \ref{fig12}; and moving
breathers (not shown here in detail), which lose the initial vorticity. In
addition, there is a parameter region in which the double vortex splits into
$2$, $3$, or $4$ breathers (each without intrinsic vorticity), see details in
Fig. \ref{fig13}.

\section{Conclusions}

In this work, we have proposed and investigated a model for the passively
mode-locked laser, based on the graded-index nonlinear multimode fiber. The
model is based on the $(3+1)$D cGLE (complex Ginzburg-Landau equation) with
the complex cubic-quintic nonlinearity and the transverse two-dimensional
trapping potential, which represents the graded-index structure. A novelty in
comparison with previously studied models is that stable stationary and
dynamical modes are found in the absence of the spatial and temporal filtering
in the model.

The systematic numerical analysis reveals a variety of lasing regimes,
including fundamental $(3+1)$D solitons and breathers, as well as stable
solitary vortices with winding number $n=1$, that can also exhibit a splitting
instability leading to formation of two fundamental solitons. The vortices
with $n=2$ are found to be unstable against splitting into a rotating bound
state of two unitary vortices. A remarkable feature of the system is the
coexistence of between the stable fundamental solitons and vortices with
$n=1$. A variety of operation regimes reported here, and a possibility of
their implementation in practical devices is expected to assist the
development of methods to optimize and tune temporal, spatial and spectral
characteristics of short spatiotemporal pulses generated in multimode fiber lasers.

\section*{Funding}

Thailand Research Fund (grant BRG6080017); Russian Foundation for Basic
Research (17-02-00081); Israel Science Foundation (grant 1287/17).


\begin{thebibliography}{99}                                                                                               %
\bibitem {grelu}P. Grelu, N. Akhmediev, \textquotedblleft Dissipative solitons
for mode-locked lasers,\textquotedblright\ Nature Photonics \textbf{6}(2),
84--92 (2012)

\bibitem {wise1}W. Fu, L. G. Wright, P. Sidorenko, S. Backus, and F. W. Wise,
\textquotedblleft Several new directions for ultrafast fiber lasers
[Invited],\textquotedblright\ Opt. Express \textbf{26}(8), 9432-9463 (2018)

\bibitem {lush}L. G. Wright, Z. M. Ziegler, P. M. Lushnikov, and Z. Zhu,
\textquotedblleft Multimode Nonlinear Fiber Optics: Massively Parallel
Numerical Solver, Tutorial, and Outlook,\textquotedblright\ IEEE J. Sel. Top.
Quantum Electron,\textquotedblright\ \textbf{24}(3), 5100516 (2018)

\bibitem {jena}C. Jauregui, J. Limpert, and A. T\"{u}nnermann,
\textquotedblleft High-power fibre lasers,\textquotedblright\ Nature.
Photonics \textbf{7}(11), 861--867 (2013).

\bibitem {sdm}D. Richardson, J. Fini, and L. Nelson, \textquotedblleft
Space-division multiplexing in optical fibres,\textquotedblright\ Nature.
Photonics \textbf{7}(5), 354--362 (2013).

\bibitem {grin1}Z. Liu, L. G. Wright, D. N. Christodoulides, and F. W. Wise,
\textquotedblleft Kerr self-cleaning of femtosecond-pulsed beams in
graded-index multimode fiber,\textquotedblright\ Opt. Lett. \textbf{41}(16),
3675--3678 (2016).

\bibitem {grin2}K. Krupa, A. Tonello, B. M. Shalaby, M. Fabert, A.
Barth\'{e}l\'{e}my, G. Millot, S. Wabnitz, and V. Couderc, \textquotedblleft
Spatial beam self-cleaning in multimode fibres,\textquotedblright\ Nature.
Photonics \textbf{11}(4), 237--241 (2017).

\bibitem {grin_a}A. S. Ahsan, and G. P. Agrawal. \textquotedblleft
Graded-index solitons in multimode fibers,\textquotedblright\ Opt. Lett.
\textbf{43}, 3345-3348 (2018).

\bibitem {grin_b}L. G. Wright, W. H. Renninger, D. N. Christodoulides, and F.
W. Wise, \textquotedblleft Spatiotemporal dynamics of multimode optical
solitons,\textquotedblright\ Opt. Express \textbf{23}, 3492-3506 (2015).

\bibitem {grin_c}O. V. Shtyrina, M. P. Fedoruk, Y. S. Kivshar, and S. K.
Turitsyn, \textquotedblleft Coexistence of collapse and stable spatiotemporal
solitons in multimode fibers,\textquotedblright\ Phys. Rev. A \textbf{97},
013841 (2018).

\bibitem {sol1}L. G. Wright, D. N. Christodoulides, and F. W. Wise,
\textquotedblleft Controllable spatiotemporal nonlinear effects in multimode
fibres,\textquotedblright\ Nature. Photonics \textbf{9}(5), 306--310 (2015).

\bibitem {sol2}Z. Zhu, L. G. Wright, D. N. Christodoulides, and F. W. Wise,
\textquotedblleft Observation of multimode solitons in few-mode
fiber,\textquotedblright\ Opt. Lett. \textbf{41}(20), 4819--4822 (2016).

\bibitem {r31}W. H. Renninger and F. W. Wise, \textquotedblleft Optical
solitons in graded-index multimode fibres,\textquotedblright\ Nature. Commun.
\textbf{4}, 1719 (2013).

\bibitem {sol3}K. Krupa, A. Tonello, A. Barth\'{e}l\'{e}my, V. Couderc, B. M.
Shalaby, A. Bendahmane, G. Millot, and S. Wabnitz, \textquotedblleft
Observation of geometric parametric instability induced by the periodic
spatial self-imaging of multimode waves,\textquotedblright\ Phys. Rev. Lett.
\textbf{116}(18), 183901 (2016).

\bibitem {exp2}H. Qin, X. Xiao, P. Wang, and C. Yang, \textquotedblleft
Observation of soliton molecules in a spatiotemporal mode-locked multimode
fiber laser,\textquotedblright\ Opt. Lett. \textbf{43}(9), 1982-1985 (2018)

\bibitem {exp3}Y. Ding, X. Xiao, P. Wang, and C. Yang, \textquotedblleft
Multiple-soliton in spatiotemporal mode-locked multimode fiber
lasers,\textquotedblright\ Opt. Express \textbf{27}(8), 11435-11446 (2019)

\bibitem {exp4}T. Chen, Q. Zhang, Y. Zhang, and X. Li, \textquotedblleft
All-fiber passively mode-locked laser using nonlinear multimode interference
of step-index multimode fiber,\textquotedblright\ Photon. Res. \textbf{6},
1033-1039 (2018)

\bibitem {exp5}D. Mao, M. Li, Z. He, X. Cui, H. Lu, W. Zhang, H. Zhang, and J.
Zhao, Optical vortex fiber laser based on modulation of transverse modes in
two mode fiber,\textquotedblright\ APL Photonics 4(6), 060801 (2019)

\bibitem {turyt}I. S. Chekhovskoy, A. M. Rubenchik, O. V. Shtyrina, and M. P.
Fedoruk, \textquotedblleft Nonlinear combining and compression in multicore
fibers,\textquotedblright\ Phys. Rev. A \textbf{94}(4), 043848 (2016)

\bibitem {jena1}S. Minardi, F. Eilenberger, Y. V. Kartashov, Three-Dimensional
Light Bullets in Arrays of Waveguides, Phys. Rev. Lett. \textbf{105}(26),
263901 (2010)

\bibitem {jena2}F. Eilenberger, K. Prater, S. Minardi, R. Geiss, U. R\"{o}pke,
J. Kobelke, K. Schuster, H. Bartelt, S. Nolte, A. T\"{u}nnermann, and T.
Pertsch, \textquotedblleft Observation of discrete, vortex light
bullets\textquotedblright, Phys. Rev. X \textbf{3}, 041031 (2013)

\bibitem {winf}H. G. Winful and D. T. Walton, \textquotedblleft Passive mode
locking through nonlinear coupling in a dual-core fiber
laser,\textquotedblright\ Opt. Lett. \textbf{17}(23), 1688--1690 (1992).

\bibitem {kutz}J. Proctor and J. N. Kutz, \textquotedblleft Nonlinear
mode-coupling for passive mode-locking: application of waveguide arrays,
dual-core fibers, and/or fiber arrays,\textquotedblright\ Opt. Express
\textbf{13}(22), 8933--8950 (2005).

\bibitem {egg}T. F. S. B\"{u}ttner, D. D. Hudson, E. C. M\"{a}gi, A. C.
Bedoya, T. Taunay, and B. J. Eggleton, \textquotedblleft Multicore, tapered
optical fiber for nonlinear pulse reshaping and saturable
absorption,\textquotedblright\ Opt. Lett. \textbf{37}(13), 2469--2471 (2012).

\bibitem {auston}D. Auston, \textquotedblleft Transverse mode
locking,\textquotedblright\ IEEE J. Quant. Electron \textbf{4}, 420-422 (1968).

\bibitem {smith}P. W. Smith, \textquotedblleft Simultaneous phase-locking of
longitudinal and transverse laser modes,\textquotedblright\ Appl. Phys. Lett.
\textbf{13}, 235 (1968).

\bibitem {firth}A. M. Dunlop, W. J. Firth, E. M. Wright, \textquotedblleft
Master equation for spatiotemporal beam propagation and Kerr lens
mode-locking,\textquotedblright\ Opt. Commun. \textbf{138}, 211-226 (1997)

\bibitem {book}N. Akhmediev and A. Ankiewicz, eds., \textit{Dissipative
solitons}, Vol. 661 of Lecture Notes in Physics (Springer, 2005).

\bibitem {agr}S. Raghavan, and G. P. Agrawal, \textquotedblleft Spatiotemporal
solitons in inhomogeneous nonlinear media,\textquotedblright\ Opt. Commun.
\textbf{180}, 377-382 (2000).

\bibitem {bul1}N. Akhmediev, J. M. Soto-Crespo, P. Grelu, \textquotedblleft
Spatiotemporal optical solitons in nonlinear dissipative media: From
stationary light bullets to pulsating complexes,\textquotedblright\ Chaos
\textbf{17} 037112 (2007)

\bibitem {bul2}N. B. Aleksi\'{c}, V. Skarka, D. V. Timotijevi\'{c},and D.
Gauthier, \textquotedblleft Self-stabilized spatiotemporal dynamics of
dissipative light bullets generated from inputs without spherical symmetry in
three-dimensional Ginzburg-Landau systems,\textquotedblright\ Phys. Rev. A
\textbf{75}, 061802 (2007)

\bibitem {bul3}S. Chen, \textquotedblleft Analytical spinless light-bullet
solutions as attractive fixed points in the three-dimensional cubic-quintic
complex Ginzburg-Landau equation,\textquotedblright\ Phys. Rev. A \textbf{86},
033829 (2012)

\bibitem {vort1}D. Mihalache, D. Mazilu, F. Lederer, H. Leblond,
\textquotedblleft Stability limits for three-dimensional vortex solitons in
the Ginzburg-Landau equation with the cubic-quintic
nonlinearity,\textquotedblright\ Phys. Rev. A \textbf{76}, 045803 (2007)

\bibitem {vort2}D. Mihalache, D. Mazilu, F. Lederer, H. Leblond
\textquotedblleft Collisions between coaxial vortex solitons in the
three-dimensional cubic-quintic complex Ginzburg-Landau
equation,\textquotedblright\ Phys. Rev. A \textbf{77}(3) (2008)

\bibitem {vort3}B. Liu, Y.-F. Liu, X.-D. He, \textquotedblleft Impact of phase
on collision between vortex solitons in three-dimensional cubic-quintic
complex Ginzburg-Landau equation,\textquotedblright\ Opt. Express \textbf{22},
26203-26211 (2014)

\bibitem {vort4}N. A. Veretenov, S. V. Fedorov, N. N. Rosanov,
\textquotedblleft Topological Vortex and Knotted Dissipative Optical 3D
Solitons Generated by 2D Vortex Solitons,\textquotedblright\ Phys. Rev. Lett.
\textbf{119}(26), 263901 (2017)

\bibitem {vort5}B. A. Malomed, \textquotedblleft(INVITED)Vortex solitons: Old
results and new perspectives\textquotedblright, Physica D \textbf{399},
108-137 (2019)

\bibitem {May}T. Mayteevarunyoo, B. A. Malomed, and D. V. Skryabin,
\textquotedblleft One- and two-dimensional modes in the complex
Ginzburg-Landau equation with a trapping potential,\textquotedblright\ Opt.
Express \textbf{26}, 8849-8865 (2018).

\bibitem {May2}T. Mayteevarunyoo, B.A. Malomed and D. Skryabin,
\textquotedblleft Vortex modes supported by spin-orbit coupling in a laser
with saturable absorption,\textquotedblright\ New J. Phys. \textbf{20},
113019, 2018.

\bibitem {yang}J. Yang, Nonlinear Waves in Integrable and Nonintegrable
Systems (SIAM: Philadelphia, 2010).

\bibitem {exp1}L. G. Wright, D. N. Christodoulides, and F. W. Wise,
\textquotedblleft Spatiotemporal mode-locking in multimode fiber
lasers,\textquotedblright\ Science \textbf{358}(6359), 94-97 (2017).
\end{thebibliography}
\end{document}